\documentstyle{article}

\title{Phase Diagram of $S=1$ Bond-Alternating XXZ chains}
\author{Atsuhiro Kitazawa, Kiyohide Nomura and Kiyomi Okamoto \\
  {\em Department of Physics, Tokyo Institute of Technology,} \\
  {\em Oh-okayama, Meguro-ku, Tokyo 152, Japan} \\
}
\date{(\today)}

\newcommand{\vS}{\mbox{\boldmath$S$}}

\begin{document}
\maketitle

\begin{abstract}
The phase transitions between the XY, the dimer and the Haldane phases of
the spin-1 bond-alternating XXZ chain are studied
by the numerical diagonalization.
We determine the phase diagram at $T=0$ and also identify
the universality class with the level spectroscopy.
We find that exactly on $\Delta=0$ there is a Berezinskii-Kosterlitz-Thouless
transition line which separates the XY and the Haldane phase and there exists a
multicritical point of the XY, the dimer and the Haldane phases.
We discuss that the critical properties of this model is
of the 2-D Ashkin-Teller type reflecting the hidden $Z_2\times Z_2$ symmetry.
\end{abstract}

Haldane \cite{H} predicted the difference of the critical properties between
integer $S$ and half odd integer $S$ spin chains.
When $S$ is an integer, the Heisenberg spin chain has a unique disordered
ground state with the finite energy gap, while when $S$ is a half odd integer
it has no energy gap and belongs to the same universality class as
the $S=1/2$ case. Den Nijs and Rommelse \cite{dNR} argued that
the Haldane gap system is characterized by the string order parameter
which is defined by the non-local operator.
Recently the symmetry breaking is
recognized as the breakdown of the hidden $Z_{2}\times Z_{2}$ symmetry
\cite{KT3}. In this letter we study the phase diagram of the
$S=1$ anisotropic spin chain with bond-alternation
\begin{equation}
  H = \sum(1+\delta(-1)^{j})[S_{j}^{x}S_{j+1}^{x}+S_{j}^{y}S_{j+1}^{y}
    +\Delta S_{j}^{z}S_{j+1}^{z}],
\label{ham}
\end{equation}
to see this symmetry breaking.
By studying the phase diagram,
we show the difference of the topology of the phase diagrams
between $S=1$ and $S=1/2$ cases.

First we review the two dimensional (2-D) Ashkin-Teller model
which will be useful to understand
the model (\ref{ham}). The 2-D Ashkin-Teller model can be thought as
two Ising models coupled by a four-spin interaction, and
has the $Z_{2}\times Z_{2}$ symmetry.
It is known that there is the 2-D Gaussian critical line of
continuously varying critical
exponents \cite{KB}, and at one end, this line breaks up into
two 2-D Ising critical lines \cite{KNO}.
At another end of the Gaussian critical line,
it meets two Berezinskii-Kosterlitz-Thouless (BKT) critical lines
\cite{BKT}, and so there is a massless region (so called "critical fan")
in the 2-D Ashkin-Teller model.
The 2-D Gaussian critical line separates the fully ordered phase
and the fully disordered phase about the $Z_{2}\times Z_{2}$ symmetry,
and the two 2-D Ising critical lines are the boundaries of the partially
ordered phase.
According to Kohmoto {\em et al}. \cite{KNK}, the 2-D Ashkin-Teller model can
be mapped to the
$S=1/2$ XXZ chain with bond-alternation whose Hamiltonian is defined by
eq.(\ref{ham}). For this $S=1/2$ model the Gaussian critical line corresponds
to
the $\delta=0$ line ($-1/\sqrt{2}<\Delta <1$) and at the $\Delta=1$, $\delta=0$
point the bifurcation to the 2-D Ising critical lines occurs.
In finite systems, it is known that the boundary condition is important
for the eigenvalue structure of these models \cite{ABB}.
For the periodic boundary conditions, the $S=1/2$ XXZ chain is $U(1)$
symmetric and the quantum Ashkin-Teller model is $Z_{2}\times Z_{2}$,
and the correspondence of the eigenvalues for both models is not complete.
On the other hand, for open boundary conditions the eigenvalues of the
$S=1/2$ XXZ chain with $2M$ sites are exactly related to those of the
self-dual $M$-site quantum Ashkin-Teller chain.

Let us return to the $S=1$ case.
For $\delta =0$ case this model has been studied numerically
\cite{BJ,KT,KN,ST,YT} in relation to the Haldane conjecture \cite{H}.
In the region $\Delta > \Delta_{c2} = 1.16\pm 0.02$ the system is N\'eel
ordered and it
has a two-fold degenerate ground state.
When $\Delta_{c2} > \Delta > \Delta_{c1} \simeq 0.0$, the system is in
the Haldane phase, where the ground state is a unique singlet
with an energy gap and the spin correlation decays exponentially.
In the region $-1< \Delta < \Delta_{c1}$ the system is
the XY phase, characterized by the power law decay of the spin
correlation function and the gapless excitation spectrum.
For $\Delta < -1$ this system is ferromagnetic long-range ordered.
The phase transition at $\Delta_{c2}$ belongs to
the 2-D Ising universality class \cite{KN,ST}.
It is thought that the BKT transition occurs at $\Delta_{c1}$.
Numerically Botet and Jullian estimated $\Delta_{c1}\sim 0.1$,
Kubo and Takada $\Delta_{c1}\simeq 0.3$,
Sakai and Takahashi $\Delta_{c1}= -0.01\pm0.03$
and Yajima and Takahashi $\Delta_{c1}=0.069\pm0.003$.
However it is difficult to evaluate the precise value of the $\Delta_{c1}$ from
the gap behavior of finite
size systems because of the singular behavior of the excitation gap
$\sim\exp(-\mbox{const}/\sqrt{\Delta-\Delta_{c1}})$.
In fact, the finite-size scaling method may lead to false conclusions for
the BKT-type transition \cite{BM,SZ,NO}. So in this letter, we mainly discuss
the method to obtain the BKT transition points.

{}From the other point of view, the isotropic case with bond-alternation
\begin{equation}
H = \sum_{j}(1+\delta(-1)^{j})\vS_{j}\cdot\vS_{j+1}
\label{iso}
\end{equation}
has been extensively studied.
Theoretically Affleck and Haldane \cite{AFF,AFFH} studied
the Heisenberg case for the arbitrary $S$, mapping to the
$O(3)$ nonlinear $\sigma$-model, and showed that
the topological angle $\theta$ is given
by $\theta=2\pi S(1+\delta)$ and the system should be massless
when $\theta/2\pi$ is half odd integer.
Numerically several authors estimated the transition point $\delta_{c}$
between the Haldane phase and the dimer phase, and
calculated the conformal anomaly $c$ using
the density matrix renormalization group method \cite{KT2},
the Monte Carlo method \cite{Y} and the exact diagonalization \cite{Te}.
They concluded $\delta_{c}\simeq 0.25$ and $c=1$.
Recently Totsuka $et$ $al.$ \cite{Te} calculated
the critical dimensions up to the logarithmic corrections and identified the
universality class as the $k=1$ $SU(2)$ Wess-Zumino-Witten (WZW) model.

According to Kennedy and Tasaki \cite{KT3}, in the Haldane phase
the hidden $Z_{2}\times Z_{2}$ symmetry is fully broken,
in the N\'eel phase only one $Z_{2}$ symmetry is broken and in the dimer phase
there is no symmetry breaking.
The edge states with open boundary conditions reflect this symmetry.
This suggests that the transition point
at $\Delta=1$, $\delta\simeq 0.25$ is the Ashkin-Teller bifurcation point.

On the anisotropic case with bond alternation,
Singh and Gelfand \cite{SG} treated this model by series expansions.
They stated that there exists a multicritical point
of the N\'eel, the dimer, the Haldane and the XY phases.
In their phase diagram there is no
2-D Gaussian critical line, so it contradicts to the expectation that
the transition point $\Delta=1$,
$\delta\simeq 0.25$ is the Ashkin-Teller bifurcation point.

In the following we consider the XY-dimer and the XY-Haldane phase transitions.
Using the bosonization method, Schulz \cite{S} showed that
the effective Hamiltonian of this transition is described by the
sine-Gordon model:
\begin{equation}
  H =\frac{1}{2\pi}\int dx\left[
    vK(\pi\Pi)^{2}+\frac{v}{K}\left(\frac{\partial\phi}{\partial x} \right)^{2}
      \right]
    +\frac{2g}{(2\pi\alpha)^{2}}\int dx\cos\sqrt{8}\phi,
\label{con}
\end{equation}
where $\Pi$ is the momentum density conjugate to $\phi$,
\begin{equation}
  [\phi(x),\Pi(x^{'})] = i\delta(x-x^{'}),
\end{equation}
$\alpha$ is the lattice constant and $v$ is the spin wave velocity.
The dual field $\theta(x)$ is defined as $\partial_{x}\theta(x) = \pi\Pi(x)$.
Here we neglect several irrelevant terms for the BKT transition.
In the full continuum Hamiltonian there is another part which is
a continuum limit of the 2-D Ising model and
it is important for the N\'eel-dimer transition
and the N\'eel-Haldane transition.
The BKT transition occurs when the scaling dimension of the operator
$\cos\sqrt{8}\phi$ becomes 2 or the renormalized $K$ is 1.

According to Cardy \cite{C}, assuming conformal invariance  of correlation
functions,
the relation between the critical dimension and the
energy gap of the finite system with periodic boundary conditions is
\begin{equation}
  E_{n}(L) - E_{0}(L) =\frac{2\pi v x_{n}}{L}
\label{ca}
\end{equation}
where $v$ is the spin wave velocity of the model, $x_{n}$ is the scaling
dimension and $L$ is the number of the unit cell
(for the present model $L$ = $N/2$, $N$ is the number of sites).
We determine the spin wave velocity from the energy gap at the
wave number $k=2\pi/L$ and $\sum S^{z}=0$ as
\begin{equation}
  v=\lim_{L\rightarrow \infty}v(L) = \lim_{L\rightarrow \infty}
 \frac{L\Delta E(k=2\pi/L)}{2\pi}.
\end{equation}
Numerically we fitted values of $v(L)$ to the form $v(L) = v + a/L + b/L^{2}$.
The leading finite size correction of the ground state energy is
\begin{equation}
  E_{0}(L) = \epsilon_{0}L - \frac{\pi v c}{6L},
\label{ano}
\end{equation}
where $c$ is the conformal anomaly number.

Besides the operator $\cos\sqrt{8}\phi$, there are several operators
whose scaling dimensions become $2$ at the BKT transition point.
Recently one of the authors \cite{N} studied the structure of these operators
for finite size systems. He found that at the BKT transition point
$\cos\sqrt{8}\phi$ and the marginal operator
\begin{equation}
  {\cal{M}} = -K(\pi\Pi)^{2}+\frac{1}{K}\left(\frac{\partial\phi}{\partial x}
  \right)^{2}
\end{equation}
which is identical to the free part of the Lagrangian density,
should be hybridized to satisfy the orthogonality condition of
the two-point correlation functions in the fields.
This hybridization comes from the existence of $\cos\sqrt{8}\phi$
term in eq.(\ref{con}) which generates the wave function renormalization.
He showed that the ratio of
leading logarithmic corrections of the scaling dimensions
of $\cos \sqrt{8}\phi$-like, $\sin\sqrt{8}\phi$,
$\exp(\mp i\sqrt{8}\theta)$ and marginal-like operators
(in the following we define the scaling dimensions of these operators as
$x_{1}$,
$x_{2}$, $x_{3}$ and $x_{0}$ respectively) is
$2:1:-1:-1$.
Near the transition point the scaling dimensions $x_{0}$ and $x_{3}$
cross linearly.
Note that in the $XY$ phase the scaling dimension of
the marginal operator ${\cal{M}}$ remains $2$ but for other operators
the scaling dimensions vary continuously as the effective $K$ varies.
Since the scaling dimension is related to the excitation energy as
eq.(\ref{ca}),
this means that at the critical point there exists a degeneracy of excited
states which correspond to the $\exp(\mp i\sqrt{8}\theta)$ and
the marginal-like operator.
So we can determine the BKT transition point by the level crossing of
these excitations, and we can also eliminate the logarithmic corrections by
the appropriate average of the excitations. We call this method the
level spectroscopy.

In order to identify the excitations with those of the sine-Gordon model,
we use the following symmetries.
Hamiltonian (\ref{ham}) is invariant under spin rotation around the
$z$-axis, translation by two-site ($S^{z}_{i}\rightarrow S^{z}_{i+2}$)
, space inversion ($S^{z}_{i}\rightarrow S^{z}_{N-i+1}$) and spin reversal
($S^{z}_{i}\rightarrow -S^{z}_{i}$). Corresponding eigenvalues are
$\sum_{i=1}^N S^z_i$, $q=4\pi n/N$, $P=\pm 1$ and $T = \pm 1$ respectively.
The symmetry operations in the sine-Gordon model are as follows \cite{N}.
The operation to the space inversion ($P$) is
\begin{equation}
  \phi \rightarrow -\phi,\hspace{5mm}
  \theta \rightarrow \theta+\sqrt{2}\pi,\hspace{5mm} x \rightarrow -x,
\end{equation}
and the operation to the spin reverse ($T$) is
\begin{equation}
  \phi \rightarrow -\phi,\hspace{5mm}
  \theta \rightarrow -\theta+\sqrt{2}\pi.
\end{equation}
The eigenvalues of these symmetry operations are summarized in Table 1.
For finite systems the ground state is singlet
($\sum S^{z}=0, q=0, P=T=1$) in the whole region ($\Delta > -1$).
In $\sum S^{z}=0$, $q=0$ subspace, the first excited state
is $P=T=1$, the second one is $P=T=-1$
and the third one is $P=T=1$ in the massless phase.
In  $\sum S^{z}=\pm 4$, $q=0$ subspace the first excited state is
$P=1$. On the BKT line the first  $\sum S^{z}=0$, $q=0$, $P=T=1$
and $\sum S^{z}=\pm 4$, $q=0$, $P=1$ excitations degenerate.
Note that in $\sum S^{z}=\pm 1$, $P=-1$ subspace the lowest excitation states
correspond to the operators $\exp( \mp i\theta/\sqrt{2})$
in the sine-Gordon language, whose scaling dimension is $x=1/8$ ($\eta=1/4$)
at the BKT transition point.

In numerical calculations, we consider finite rings of $N$ sites
($N = 6, 8, 10, 12$) with periodic boundary conditions.
We calculate the excitation energies of $\sum S^{z}=0$ and $\sum S^{z}=\pm4$
corresponding to $\cos \sqrt{8}\phi$, $\sin\sqrt{8}\phi$,
marginal operator and $\exp(\mp i\sqrt{8}\theta)$.
In Fig.1(a) we show the excitation energies for $N=12$, $\Delta=-0.5$.
It is seen from this figure that the marginal-like excitation
($\sum S^{z} =0$) and the
excitation corresponding to the operator $\exp(\mp i\sqrt{8}\theta)$
($\sum S^{z} =\pm4$) cross
linearly. We determine the critical point of the XY-dimer phase transition by
the crossing point of these excitations as mentioned above.
Figure 1(b) shows the excitation energies for $N=12$, $\Delta=0$.
In this case one of the state ($\sum S^{z}=0$, $q=0$, $P=T=1$)
and the eigenstate ($\sum S^{z}=\pm4$, $q=0$, $P=1$) are degenerate
in the whole region (numerically this degeneracy is exact even for finite
systems),
thus we conclude that there is a BKT transition line between the XY and the
Haldane phases on $\Delta=0$.
Finally we obtain the multicritical point of the XY, the dimer and the
Haldane phases by the crossing
of the excitations corresponding to the operators
$\sin\sqrt{8}\phi$ $(\sum S^{z}=0, P=-1)$ and
$\exp(\mp i\sqrt{8}\theta)$ $(\sum S^{z}=\pm4, P=1)$.
This multicritical point corresponds to
$K=1$ and $g=0$ in the continuum Hamiltonian (\ref{con}).
We estimate this point as $\delta=0.22\pm 0.01$.
Figure 2 indicates the full phase diagram of this model.
Other boundaries (between the dimer, the Haldane and the N\'eel phases)
are determined by the usual finite-size scaling method.

In order to verify that the XY-dimer and the XY-Haldane phase transitions
are of the BKT-type, we present
$(x_{0}+x_{2})/2$ and $(2x_{0}+x_{1})/3$ in Fig. 3 which eliminate
the logarithmic corrections of the scaling dimension.
Since we only investigate small size systems, these values are not
exactly equivalent to $2$, due to the existence of irrelevant fields,
but they converge rapidly to this value with
increasing system size.
Also we estimate the conformal anomaly using eq.(\ref{ano}), as
$c=0.97$ for $\Delta=-0.5$ and $c=0.98$ for $\Delta=0$, $\delta=0$.
Hence we can conclude that the transitions are of the BKT-type.
On the other hand the phase transition between the Haldane and the dimer phases
is thought to belong to the 2-D Gaussian universality class. In this
case the operator $\cos\sqrt{8}\phi$ in the effective Hamiltonian
eq.(\ref{con}) is relevant, that is, the effective $K$ is smaller
than $1$ and the second order transition occurs at $g=0$.

Now we consider the reason why the boundary of the XY and the Haldane phases is
just on the $\Delta=0$ line.
On this line $\sum S^{z}=0$ and $\sum S^{z}=\pm4$ excitations are degenerate
exactly.
Note that in the case of the open boundary conditions, similar degeneracy was
reported by Alcaraz and Moreo \cite{AM}, who discussed that it strongly
suggests
$\Delta_{c1}=0$.
Although we have no explicit explanation on these degeneracies,
we remark the following symmetry.
By the unitary operator $U=\exp[i\pi\sum_{j=1}^{N}(j+1)S_{j}^{z}]$,
the Hamiltonian is transformed as
$UH(\Delta)U^{-1} = -H(-\Delta)$.
This means that there is an additional symmetry at $\Delta=0$.
When $\Delta=0$, for an eigenstate $|\psi_{0}\rangle$, $U|\psi_{0}\rangle$
is also an eigenstate of $H$.
Perhaps this symmetry makes the phase boundary of the XY and the
Haldane phases on the line $\Delta = 0$ for the $S=1$ case.
Relating to this, Alcaraz and Hatsugai \cite{AlH} showed that
the $S=1$ XY model does not have the string order,
and their argument can also be applied to the bond-alternation cases.
In comparison, the $S=1/2$ XY model with bond-alternation is equivalent to
the free fermion model. In this case, there exists the degeneracy between
$\sum S^{z}=0$ and $\sum S^{z}=\pm 2$,
and the second order transition occurs at $\delta=0$.

Let us discuss the relation between the present phase diagram and the
bosonization by Schulz. The renormalization flow of eq.(\ref{con}) is
shown in Fig.4. Our phase diagram Fig.2 means that all the starting points
of the renormalization for $\Delta=0$ and $0<\delta<\delta_{c}$ lie just on
the line in Fig.4 which separates the XY and the Haldane phases,
and the multicritical point of the XY, dimer, Haldane phases ($\Delta=0$,
$\delta=\delta_{c}$) corresponds to the origin of Fig.4. Although Schulz
stated that the bare values of $K$ and $g$ are $K=\sqrt{1+(2\Delta/\pi)}$
(independent of $\delta$) and $g=-\Delta+\delta^{2}$ respectively,
it is easy to see that his expression cannot explain the above-mentioned
behavior.
We consider that this discrepancy comes from that Schulz dropped many
operators in the course of the bosonization.
However, apart from the bosonization, the BKT multicritical structure can be
expected based on the $U(1)\times Z_{2} \times Z_{2}$ symmetry \cite{N}.
If the bare values have the form, for instance,
$K=\sqrt{1+(2\Delta/\pi)-a_{1}\delta^{2}}$ and $g=a_{2}-\Delta+a_{3}\delta^{2}$
with positive constants $a_{1}$, $a_{2}$ and $a_{3}$, the renormalization
behavior of eq.(\ref{con}) is consistent with our phase diagram Fig.2.
Unfortunately
we could not derive such expressions from the original spin
Hamiltonian. To derive the correct form of the effective Hamiltonian is a
future problem.

Lastly, we argue the difference of the phase diagram between the $S=1/2$ and
the $S=1$ cases.
Note that the phase diagram is symmetric on the transformation
$\delta\rightarrow-\delta$.
Reflecting the hidden $Z_{2}\times Z_{2}$ symmetry,
the $S=1/2$ model and the $S=1$ model with bond alternation have the same
phase diagram with the 2-D Ashkin-Teller model.
For the $S=1/2$ model,
there are two BKT lines, one 2-D Gaussian critical line and two 2-D Ising type
critical lines for $-1<\delta<1$. On the other hand, the $S=1$ model
has three BKT lines, two 2-D Gaussian
lines and three 2-D Ising type critical lines.
Oshikawa \cite{OSHI} studied the hidden $Z_{2}\times Z_{2}$ symmetry of the
arbitrary integer $S$ Heisenberg Hamiltonian with bond-alternation
(\ref{iso}). In the content of the VBS picture, we expect
the successive dimerization transition which is in agreement with
the Affleck-Haldane prediction. From these considerations, we can expect that
for the arbitrary $S$ XXZ chain with bond-alternation, there are
$2S+1$ BKT lines, $2S$ 2-D Gaussian lines and $2S+1$ 2-D Ising lines.
In addition, for half odd integer $S$,
there is a Gaussian line in the region $\delta=0$,
$-1<\Delta<1$ and at $\Delta = 1$ the model becomes the $k=1$ $SU(2)$ WZW
model.

In summary, we present the phase diagram of the
$S=1$ bond-alternating XXZ chain (Fig.2)
which is determined by the level crossing of some special excitations on the
basis
of renormalization group analysis (Fig.1)
and also show the logarithmic corrections of the critical dimensions
on the BKT critical line (Fig.3).
There exists a BKT line which separates the XY and the Haldane
phases on $\Delta = 0$ line where the $\sum S^{z} =0$ and
the $\sum S^{z} =\pm4$ eigenstates are degenerate on the whole line.
The critical properties are recognized as that of the 2-D Ashkin-Teller
model. More detailed studies of this model will be presented in elsewhere.

K. N. acknowledge Professor H. J. Schulz for useful discussions
on the bosonization of the $S=1$ model.
The authors thank to Dr. M. Yamanaka for discussions on the Ashkin-Teller
model.
This work is partially
supported by Grant-in-Aid for Scientific Research (C) No.06640501
from the Ministry of Education, Science and Culture, Japan.

\begin{table}[h]
\begin{tabular}{cccc|c|c}\hline
\multicolumn{4}{c|}{Symmetry operation} & &  \\
 $q$ & $\sum S^{z}$ & $T$ & $P$ &
\multicolumn{1}{c|}{\raisebox{1.5ex}[0pt]{
    \begin{tabular}{c}
      Operators of \\
      sine-Gordon model
    \end{tabular}
}} &
\multicolumn{1}{c}{\raisebox{1.5ex}[0pt]{
    \begin{tabular}{c}
    Notation of \\
    scaling dimensions
\end{tabular}
}}\\
\hline
  $0$ & $0$ & $1$ & $1$ & ${\cal{M}}$ & $x_{0}$  \\
  $0$ & $0$ & $1$ & $1$ & $\cos\sqrt{8}\phi$ & $x_{1}$ \\
  $0$ & $0$ & $-1$ & $-1$ & $\sin\sqrt{8}\phi$ & $x_{2}$ \\
  $0$ & $\pm4$ & * & $1$ & $\exp\mp i\sqrt{8}\theta$ & $x_{3}$ \\ \hline
\end{tabular}
\caption{Eigenvalues of symmetry operations}
\end{table}
\clearpage

\noindent
Figure Caption

\noindent
{\bf Fig.1}
Excitation energies with $q=0$,$\sum S^{z} = 0$
and $\sum S^{z} = \pm4$ for (a) $N=12$, $\Delta=-0.5$
and (b) $N=12$, $\Delta=0$.
$\Diamond$'s are $\sum S^{z}=0$, $q=0$, $P=1$, $T=1$ excitations,
$\Box$ is $\sum S^{z}=0$, $q=0$, $P=-1$, $T=-1$ excitation
and $+$ is the $\sum S^{z}=\pm4$, $q=0$, $P=1$ excitation.

\noindent
{\bf Fig.2}
The phase diagram in the $\Delta$-$\delta$ plane.
F means the ferromagnetic long-range ordered region.
The XY-dimer and the XY-Haldane phase boundaries are of the BKT type,
the dimer-Haldane boundary is of the 2-D Gaussian type and the N\'eel
phase boundaries are of the 2-D Ising type.

\noindent
{\bf Fig.3}
Scaling dimensions eliminating logarithmic corrections
$(x_{1}+x_{3})/2$ ($+$) and $(x_{2}+2x_{3})/3$ ($\Box$) at the BKT transiton
points. (a) plots on the BKT line between the XY and the
dimer phases and (b) plots on the BKT line between
the XY and the Haldane phases ($\Delta=0$).
Bare values are also presented ($\Diamond$).

\noindent
{\bf Fig.4}
The renormalization flow of eq.(\ref{con}).

\end{document}